\documentclass[prl,twocolumn]{revtex4}
\usepackage{epsfig,amsfonts,graphicx}

\begin{document}

\title{Anomalous state of a 2DEG in vicinal Si MOSFET in high magnetic fields}
\author{Z.D. Kvon$^1$, Y.Y. Proskuryakov$^2$, A.K. Savchenko$^2$}

\address {$^1$ Institute of Semiconductor Physics, Novosibirsk, 630090 ,
Russia\\
$^2$ School of Physics, University of Exeter, Stocker Road, Exeter, EX4 4QL, U.K.
 \\}

\begin{abstract}
We report the observation of an anomalous state of a 2D electron gas near a vicinal surface of a silicon MOSFET in high
magnetic fields. It is characterised by unusual behaviour of the conductivities $\sigma _{xx}$ and $\sigma _{xy}$, which
can be described as a collapse of the Zeeman spin splitting accompanied by a large peak in $\sigma _{xx}$ and an anomalous
peak in $ \sigma _{xy}$. It occurs at densities corresponding to the position of the Fermi level above the onset of the
superlattice mini-gap inherent to the vicinal system. The range of fields and densities where this effect exists has been
determined, and it has been shown that it is suppressed by parallel magnetic fields.
\end{abstract}


\maketitle

More than twenty years after the discoveries of the QHE and FQHE,
the properties of two-dimensional electron systems remain one of
the most important subjects in condensed matter physics
\cite{1,2}, with much attention paid to the problem of
electron-electron interactions. Examples include recent
experiments indicating the formation of the quantum Hall
ferromagnetic states in double 2D layers \cite{3} and in
multi-valley 2D systems \cite{4}. The object of our study is a
high-mobility 2DEG near a vicinal surface - one that is tilted by
a small angle with respect to the widely studied (100)Si. It is
known that such tilting creates a superlattice on the Si surface
which enhances the inter-valley interaction \cite{5,6}. The effect
of this enhancement is a subject of our study. An indication of
the importance of increased valley interaction has been found in
earlier works where the anomalously large value of the
superlattice minigap \cite{5,6} and anomalies in the $\rho _{xx}$
and $\rho _{xy}$ in strong magnetic fields \cite{7,8} have been
observed.

We focus our study on properties of a 2DEG near a vicinal silicon
surface in strong perpendicular and parallel magnetic fields, and
compare its behaviour with that of a 2DEG on (100)Si. We have
observed strong anomalies in the magneto-transport of the vicinal
system which are characterised by a collapse of the Zeeman
splitting. It occurs at electron densities close to the
superlattice minigap and persists at higher densities.  We suggest
that the primary reason for this behaviour is the interplay
between enhanced inter-valley and electron-electron interactions.

The studied samples are MOSFET Hall bars on vicinal and (100)
silicon surfaces fabricated on the basis of the same technology.
The vicinal surface is tilted by the angle $9^o$ $40^{\prime }$ to
(100)Si around the direction [011]. This surface has a
superlattice structure with a period of $ 1.62$ nm. The peak
mobility is about $25000$ cm$^2$/Vs at $T=1$ K. Measurements have
been carried out in a standard four-terminal arrangement with
currents $I=1-10$ nA, in magnetic fields up to 15 T and in the
temperature range from 50 mK to 4.2 K. All results on the vicinal
sample have been qualitatively the same for the two orientations
of the Hall bar - along and perpendicular to the superlattice
axis, and below we will only discuss the latter case.

\begin{figure}[t]
\begin{center}
\includegraphics*[totalheight=4.75in]{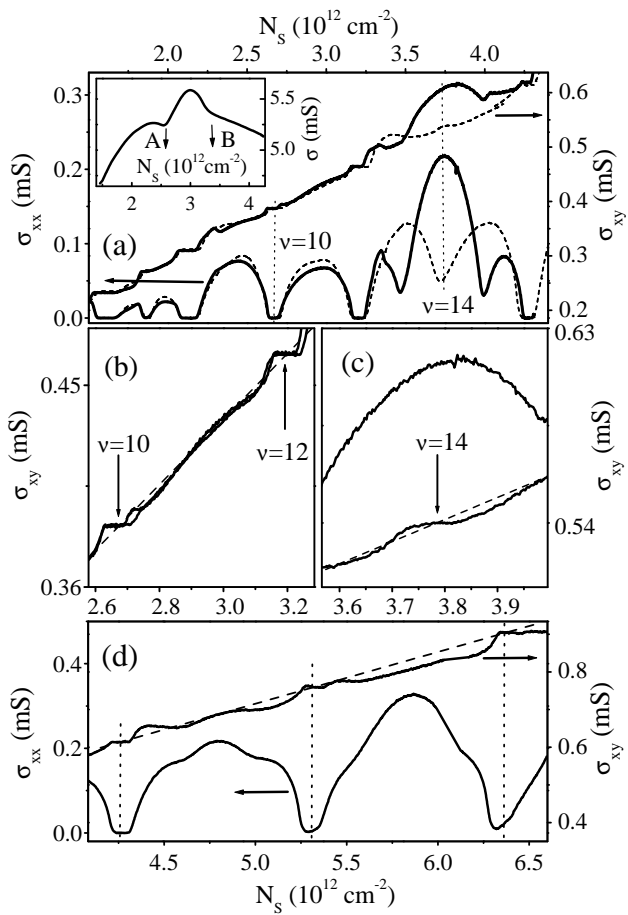}
\caption{(a) $\sigma _{xx}(N_s)$ and $\sigma _{xy}(N_s)$ at $B=11$ T and $%
T\sim 50$ mK for vicinal (solid lines) and (100)Si (dashed lines)
samples. (b, c) $\sigma _{xy}(N_s)$ for both samples zoomed-in at
low ($N_s<N_c$) and high ($N_s>N_c$) densities. (d) $\sigma
_{xx}(N_s)$ and $\sigma _{xy}(N_s)$ for the (100) sample continued
to larger $N_s$ compared with (a). Dashed lines in (b, c, d) show
the classical Hall conductivity $eN_s/B$. Inset: density
dependence of $\sigma _{xx}$ at $B=0$ and $T\simeq 50$ mK (points
A and B correspond to the boundaries of the minigap).}
\end{center}
\end{figure}

The inset to Fig. 1 shows the conductivity $\sigma_{xx} $ versus
electron density at zero magnetic field and $T\sim 50$ mK. A well
known ``$\Omega $ feature" \cite{5,6} appears at electron density
$N_s^\Delta =2.6\times 10^{12}$ cm$^{-2}$ due to the superlattice
minigap. The longitudinal conductivity $\sigma _{xx}$ and Hall
conductivity $\sigma _{xy}$ in high magnetic field $B=11$ T are
plotted in Fig. 1 (a) as functions of electron density $N_s$. A
change in $ \sigma _{xx}$ is clearly seen when the filling factor
$\nu$ increases from 10 to 14: instead of the spin-splitting dip
corresponding to $\nu =14$, a large peak appears at this filling
factor. The peak in $ \sigma _{xx}$ is accompanied by a peak in
$\sigma _{xy}(N_s)$, instead of the usual quantisation plateau.
The magnitude of the peak in $\sigma _{xy}$ significantly exceeds
the classical value: $\sigma _{xy}>eN_s/B$, Fig. 1 (c). Note that
the ordinary (100)Si MOSFET does not show such features around
$\nu =14$, where the ordinary spin splitting minimum in $\sigma
_{xx}$ and plateau in $\sigma _{xy}$ are seen. Importantly, both
systems have identical behaviour up to the filling factor $\nu
=12$, with spin splitting minima in $\sigma _{xx}$ and
corresponding plateaus in $\sigma _{xy}$.

We would like to emphasise that the anomalies in the vicinal 2DEG
are essentially different from the triple `cusp'-structure
\cite{5} seen at large densities ($N_s \sim 5\times10^{12}$
cm$^{-2}$)  in ordinary 2DEGs on (100)Si, Fig. 1(d).
\textit{Firstly}, the cusp feature in the middle of the Landau
level is a weak triple feature. In the vicinal sample, however,
the three peaks are well resolved and the central peak is more
than two times larger than the side peaks. \textit{Secondly,} the
cusp feature in (100)Si MOSFETs is not accompanied by a peak in
Hall conductivity, Fig. 1(d).
\begin{figure}[t]
\begin{center}
\includegraphics*[totalheight=4.0in]{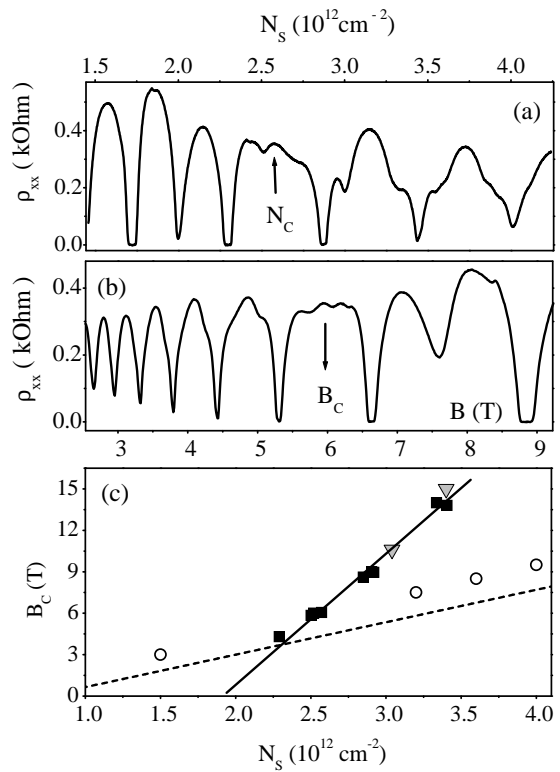}
\caption{(a) $\rho _{xx}(N_s)$ at $B=6$ T. (b) $\rho _{xx}(B)$ at $%
N_s=2.57\times 10^{12}$ cm$^{-2}$, $T\simeq 50$ mK. (c) $B_c(N_s)$
diagram. Black symbols - from $\rho _{xx}(N_s,B)$ data in
perpendicular magnetic field, with the solid line as a linear fit.
Grey symbols - from the data in tilted magnetic field. The dashed
line and open circles are, respectively, the theoretical
dependence \cite{5} and experimental data \cite{5,10.5} for
(100)Si.}
\end{center}
\end{figure}

For further consideration it is convenient to present the data in
terms of the directly measured longitudinal resistivity $\rho
_{xx}$, which qualitatively has the same character as the
conductivity $\sigma_{xx}$. Fig. 2(a) shows the $N_s$-dependence
of $\rho _{xx}$ at $B=6$ T where we study in detail how the
anomalous state appears with increasing density. At $N_s<N_c$ the
curve $\rho _{xx}(N_s)$ exhibits conventional deep minima
corresponding to cyclotron- and spin-splitting gaps. At
$N_s\approx N_c$, the behaviour of $ \rho _{xx}(N_s)$
significantly changes. The deep spin-splitting minimum disappears
and a weak structure appears in the middle of the Landau level
(LL) (later we will refer to this shape of $\rho _{xx}$ as the
`intermediate' state). Importantly, the intermediate state
corresponds to $N_c \approx N_s^\Delta $. As the density increases
further, a large central maximum develops accompanied by two side
minima that are very similar in their position to the usual
valley-splitting minima.

A similar transition from the `normal' to `anomalous' state is
seen with decreasing magnetic field at fixed electron density $N_s
= 2.57\times10^{12}$ cm$^{-2}$ close to both $N_{s}^{\Delta}$ and
$N_c$, Fig. 2(b).  Magnetic field $B_c$ corresponds here to the
`intermediate' state, therefore the appearance of the
`intermediate' state is determined by both the electron density
and magnetic field. In Fig. 2 (c) we plot the values of $N_c$ and
$B_c$, obtained as described above (black symbols). The resulting
relation is linear and can be treated as the boundary between two
`phases': one is a spin-split state, and the other is a
spin-collapsed state. This diagram allows us to come to the
following conclusions: i) the anomalous state exists only at
electron densities exceeding some density
$N_c\simeq2\times10^{12}$ cm$^{-2}$ $\sim N_s^{\Delta} $; ii) a
small increase of $ N_s$ leads to a significant increase of the
critical magnetic field $B_c$.

Fig. 3 (a, b) shows the evolution of the spin-split state into the
anomalous state around $\nu=14$ in the fourth LL as $N_s$ is
varied by small increments and the Landau level shifts to higher
fields and changes its shape. There is an interesting fact in Fig.
3 (a, b) which is not reflected in the phase diagram in Fig. 2
(c): the transition from the normal (spin-split) into the
intermediate state requires a significantly larger change of the
electron density compared with the transition from the
intermediate to the anomalous state. (In the case presented in
Fig. 3 (a, b) the first transition requires $\Delta N_s\sim40\%$,
while the second occurs with $\Delta N_s\sim5\%$.) This means that
the anomalous state develops rapidly when the Fermi level is
lifted in the minigap.
\begin{figure}[b]
\begin{center}
\includegraphics*[totalheight=4.2in]{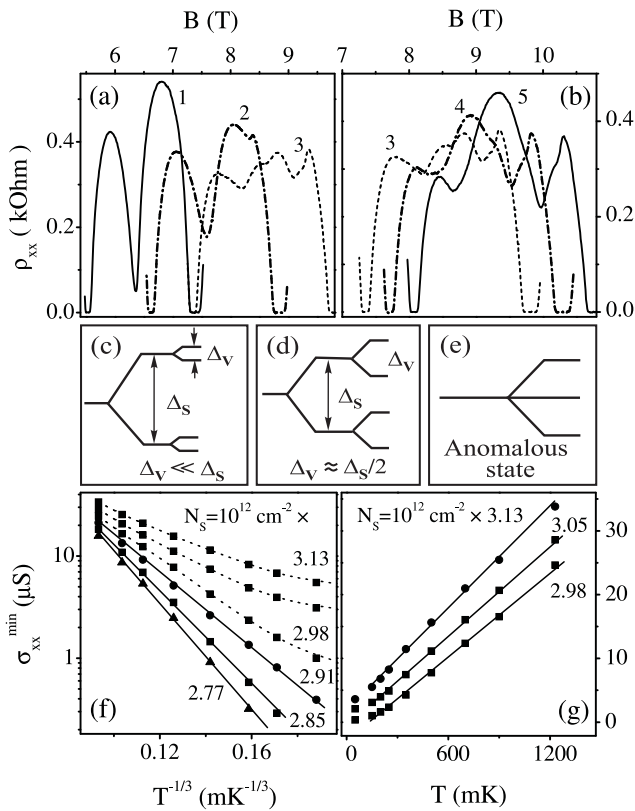}
\caption{(a) $\rho_{xx}(B)$ for Landau level $N = 4$ at different
$N_s$ in the transition from a spin-split state to an intermediate
state at $\nu=14$ (curve ``1" - $N_s = 2.15\times10^{12}$
cm$^{-2}$, ``2" - $N_s = 2.57\times10^{12}$ cm$^{-2}$, ``3" - $N_s
= 2.85\times10^{12}$ cm$^{-2}$). (b) Transition from the
intermediate state to a spin-collapsed state (``3" - $N_s =
2.85\times10^{12}$ cm$ ^{-2}$, ``4" - $N_s = 2.98\times10^{12}$
cm$^{-2}$, ``5" - $N_s = 3.13\times10^{12}$ cm$^{-2}$). (c, d, e)
The energy diagrams illustrating the transformation of LL with
increasing valley splitting. (f, g) Temperature dependence of $
\sigma_{xx} $ at the minimum $\nu=10$ at different densities, in
the transition from state (c) to state (d), see text.}
\end{center}
\end{figure}
To obtain more detailed information on how spin-splitting
disappears, the temperature dependence of $\sigma _{xx}$ was
measured. In Fig. 3 (f) we present the results for minima at $\nu
=10$. It is seen that as $N_s$ increases, it is not only the depth
of the minimum $\sigma _{xx}^{min}$ that becomes smaller, but also
the character of the temperature dependence changes. For the
deepest minimum it is the dependence $\sigma _{xx}^{min}=A\exp
[-(T_0/T)^{x}]$, with $x$ close to $1/2$ or $1/3$ (comparison with
$1/3$ is shown in Fig. 3(f) and the same agreement is obtained for
$x=1/2$). This behaviour corresponds to variable-range hopping via
localised states often observed in the QHE regime \cite{9,9.5}. As
the density is increased and the minimum becomes less pronounced,
the temperature dependence becomes linear with saturation at
lowest $T$, Fig. 3(g), which indicates a change of electron
transport from hopping to metallic. Note that this change in the
character of transport occurs before the appearance of the
intermediate state, when the spin-splitting in SdH oscillations is
still clearly seen.

Since the intermediate state has a characteristic four-fold
feature with small oscillations of equal magnitude, we assume that
it corresponds to the situation with the following relation
between valley- and spin-splitting: $\Delta_v\approx\Delta_s/2$
(see Fig. 3c). Using the theoretical dependence $\Delta_v(N_s)$
from \cite{5}, in Fig. 2 (c) we plot the corresponding $B_c(N_s)$
required to provide the equality $\Delta_v\approx\Delta_s/2$ in
the ordinary (100) structures (also the experimental points from
\cite{5,10.5} obtained for (100) samples) and compare it with the
$B_c(N_s)$ dependence in our vicinal sample. The $B_c(N_s)$ in
(100)Si starts from the origin and has a significantly smaller
slope, while the experimental diagram $B_c(N_s)$ has a distinct
threshold character. The two dependences intersect at
$N_c\simeq2.3\times10^{12}$ cm$^{-2}$ -- the value which is close
to $N_s^{\Delta}$. This clearly indicates the relation between the
anomalous state in the vicinal sample and the presence of the
minigap in it.
\begin{figure}[t]
\begin{center}
\includegraphics*[totalheight=2.7in]{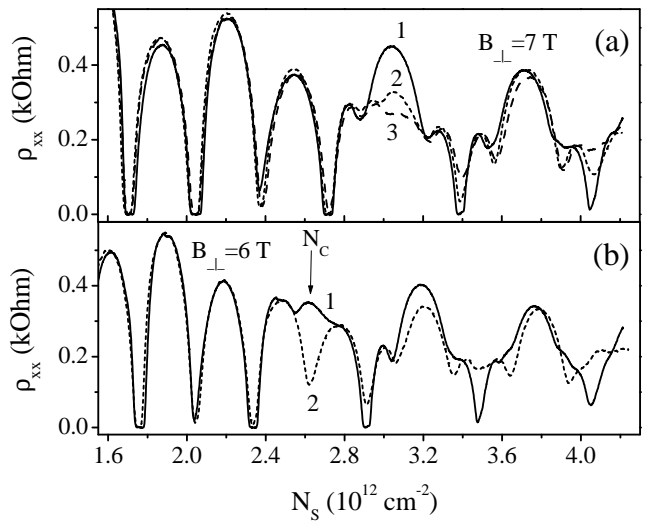}
\caption{Transformation of $\rho_{xx}(N_s)$ with increasing
parallel magnetic field $B_{||}$: (a) at perpendicular field
$B_{\perp}=7$ T, curve ``1" for $ B_{||}=0$ T, ``2" for
$B_{||}=5.87$ T, ``3" for $B_{||}=8$ T ; (b) at $B_{\perp}=6$ T,
curve ``1" for $B_{||}=0$ T, ``2" for $B_{||}=6.9$ T.}
\end{center}
\end{figure}

To prove the spin-related nature of the anomalous state, we have
studied the effect of a parallel field on it. Fig. 4(a) shows the
results of measurements of $\rho _{xx}(N_s)$ in a fixed
perpendicular magnetic field, $B=7$ T, with the parallel component
$ B_{||}$ changing gradually from $0$ to $8$ T. It is clearly seen
that the parallel field destroys the anomalous state for a
particular LL with $N=5$ and transforms it into the intermediate
state. The fact that parallel field affects only one LL can be
easily understood in terms of the ``phase'' diagram in Fig. 2(c).
The position of the Landau level with the triplet shape (marked by
``1'') corresponds to the `anomalous' side of the diagram.
Applying the parallel field shifts the state up in the vertical
direction until the boundary is reached, ``3''. (Plotting the
position of the state ``3" on the diagram, with the total magnetic
field $B_c=10.6$ T, $N_c=3.05\times 10^{10}$ cm$^{-2}$, shows that
it falls exactly on the $B_c(N_s)$ line.) Clearly, other LLs in
Fig. 4 (a) are far away from the boundary line, and hence they
cannot be moved by parallel field into the intermediate state. A
similar result has been obtained at a perpendicular field of 10 T
and total magnetic field 15 T. Both cases are plotted in Fig. 2(c)
by grey symbols. The transition for one of the Landau levels,
$N=5$, now from the intermediate to the normal spin-split state,
is shown in Fig. 4 (b). Here the parallel component of magnetic
field is increased from 0 to 6.9 T at a constant perpendicular
field of 6 T.

Let us now discuss possible origins of the anomalous state.
Spin-splitting collapse, in the presence of disorder, was
considered for a 2DEG in \cite{11}. This theory, however, cannot
explain the main features of the phenomenon we have observed. In
particular, it does not predict the presence of the triplet-like
shape of $\sigma_{xx}$. It also does not explain the change of the
transport mechanism when the the spin-split minimum is still
pronounced, Fig. 3 (f, g).

For further consideration we make a suggestion that the
dissipative conductivity is determined by the density of states
($\sigma _{xx}\propto D(E_F)$). Then the presence of the
triplet-like feature in $\sigma _{xx}$ (or $\rho _{xx}$) allows
one to speculate about the reconstruction of the spin states in
the presence of strong inter-valley interaction. At $N_s>N_c$ the
electron state originating from mixing of two interacting valley
states with different spins, can be constructed as a singlet-like
and triplet-like spin states. Then the central peak of $\sigma
_{xx}$ corresponds to the superposed singlet and the state $S_z=0$
of the triplet. Zero spin explains the absence of spin-splitting
in this case. Also, this explains the larger magnitude of the
central peak compared with the side peaks - this peak corresponds
to the state with double degeneracy (two states with zero spin),
while the side peaks represent single states with $S_z=\pm 1$. The
energy diagram of the transformation of a Landau level with
increasing density (increasing the valley splitting) is shown in
Fig. 3 (c, d, e) and corresponds respectively to the curves ``1",
``3" and ``5" in Fig. 3 (a, b).

There is an important result, which has not been discussed above.
This is the observation of the peak in Hall conductivity in the
center of the LL (Fig 1 b), where $\sigma _{xy}$ exceeds its
classical value $eN_s/B$. According to the conventional theories
of 2DES in high magnetic fields (from the early theories \cite{5}
to the modern theories of integral QHE \cite{11.5,12,13,14}), at
the maximum of $\sigma _{xx}$ the Hall conductivity always obeys
the following condition: $\sigma _{xy}\leq eN_s/B$. To date
numerous experiments \cite{12} for different 2D electron systems
including (100) Si-MOSFETs \cite{5,15} (see also Fig. 1 d)
supported this relation. It also remains valid in the FQHE regime
\cite{13,14,16}, that is in the case of strongly interacting
electrons. The peak in Hall conductivity exceeding $eN_s/B$ and
accompanying the peak in longitudinal conductivity indicates that
in the anomalous state this fundamental relation between $\sigma
_{xx}$ and $\sigma _{xy}$ breaks down. Presumably this is a
signature of some collective state resulting from the interplay
between electron-electron and intervalley interaction enhanced in
vicinal Si-MOSFETs. One can naively speculate about an enhancement
of the effective charge $e^*$ of the quasiparticles in this state,
which is larger than the free electron charge ($e^*> e$). However,
such a simplified assumption has to be proved theoretically, and
we believe that a description of the observed anomalies is
important and challenging task for the theory of correlated
multi-valley electron systems.

In summary, we have performed a detailed investigation of an
unusual state of a 2D electron gas near Si vicinal surface. This
state is characterised by large peak in $\sigma _{xx}$ and
anomalous peak in $\sigma _{xy} $, and can be described as
collapse of the Zeeman spin-splitting at electron densities when
the Fermi level enters the superlattice minigap. We have shown
that the transition to the spin-collapsed state is dependent on
electron density and magnetic field and have found the boundary
$B_c(N_s)$ between the normal and anomalous states.  We suggest
that the anomalous state is caused by the electron-electron
interaction in the condition of enhanced inter-valley interaction
in the 2DEG on the vicinal surface.

We are grateful to D. G. Polyakov and B. I. Shklovskii for reading
the manuscript and valuable comments. This work was supported by
the Royal Society, RFBR (grant 02-02-16516), Russian Ministry of
Education (program "Integration") and RAS (program "Low
dimensional quantum structures").

\end{document}